\documentclass[10pt,tightenlines]{revtex4-1}
\usepackage[dvips]{graphicx}
\usepackage{amssymb,amsfonts,amsmath}
\usepackage[usenames]{color}
\usepackage[british]{babel}

\begin{document}

\title{Evidences of two distinct local structures of water from ambient to supercooled conditions by optical Kerr effect experiments}

\author{A. Taschin$^{1}$, P. Bartolini$^{1}$, R. Eramo$^{1,2}$, R. Righini $^{1,3}$, and R. Torre$^{1,4}$}\email{corresponding author: torre@lens.unifi.it}

\affiliation{
 $^1$European Lab. for Non-Linear Spectroscopy (LENS), Univ. di Firenze, via N. Carrara 1, I-50019 Sesto Fiorentino, Firenze, Italy.\\
 $^2$Istituto Nazionale Ottica, CNR, Largo Fermi 6, I-50125 Firenze, Italy.\\
 $^3$Dip. di Chimica, Univ. di Firenze, via Della Lastruccia 13, I-50019 Sesto Fiorentino, Firenze, Italy.\\
 $^4$Dip. di Fisica e Astronomia, Univ. di Firenze, via G. Sansone 1, I-50019 Sesto Fiorentino, Firenze, Italy.}

\date{\today}

\begin{abstract} 
The liquid and supercooled states of water show a series of anomalies whose nature is lively debated. A key role is attributed to the formation of structural aggregates induced by critical phenomena occurring deep in the supercooled region; the nature of the water anomalies and of the hidden critical processes remains elusive.
Here we report a time-resolved optical Kerr effect investigation of the vibrational dynamics and relaxation processes in supercooled bulk water. The experiment measures the water intermolecular vibrations and the structural relaxation process in an extended temperature range, and with unprecedented data quality. A mode-coupling analysis of the experimental data enables to characterize the intermolecular vibrational modes and their interplay with the structural relaxation process. The results bring evidence of the coexistence of two local configurations, which are interpreted as high density and low density water forms, with an increasing weight of the latter at low temperatures. 

Published in:

\textbf{Nature Communication vol.4, pag.2401, 2013; DOI: 10.1038/ncomms3401}.
\end{abstract}

\maketitle

\section*{Introduction}
Water can remain liquid below its melting point and stays in its metastable phase known as supercooled water~\cite{mishima_98,debenedetti_03}. Several thermodynamic (e.g. isothermal compressibility, isobaric heat capacity and thermal expansion coefficient) and dynamic (e.g. viscosity, diffusivity, structural relaxation) observables of liquid water show an anomalous behaviour~\cite{debenedetti_03b}. In the supercooled state the values of these physical observables clearly diverge, with a singularity temperature $T_\mathrm{s}$ estimated at $\simeq 228$ K. Despite the wide interest and the numerous research efforts on this subject, the origin of these critical phenomena remained unclear~\cite{soper_08,angell_08,nilsoon_11} and lively debated~\cite{moore_09,moore_11,limmer_11,sciortino_11,liu_12,kesselring_12,gallo_12,poole_13}.

It is widely accepted that the water anomalies are connected with the formation of various structural aggregates whose features are modified by the temperature/pressure conditions~\cite{debenedetti_03b}. A crucial point is the physical origin of these structural aggregations. 

In their pioneering work Poole, Sciortino, Essmann, and Stanley~\cite{poole_92} postulated the existence of a liquid-liquid critical point (LLCP) based on a simulation study and on the experimental observation of polyamorphism in glassy water (i.e. the existence of two forms of water glass: low-density and high-density amorphous ice). In the LLCP hypothesis a first order phase transition takes place involving two supercooled water liquid forms: the low-density and the high-density phases. The low-density phase is characterized by tetrahedral intermolecular coordination, while the high-density one shows more packed structures with distorted networks. The existence of a LLCP is compatible with a series of computational and experimental results~\cite{debenedetti_03b}. Recent new simulation studies confirmed the LLCP hyphotesis~\cite{sciortino_11,liu_12,kesselring_12,gallo_12,poole_13}.
Nevertheless, the matter is still under debate: in their recent simulation study, Limmer and Chandler interpreted the water anomalies as non-equilibrium phenomena associated to the liquid-crystal transition~\cite{limmer_11}. Moreover, the relevance of ice nucleation phenomena and their relationship with the water anomalies has been reported~\cite{moore_09,moore_11}. 

A fundamental question in order to explain the origin of water anomalies is if two distinct intermolecular local and transient configurations, whose densities are different, persist at temperature higher than $T_\mathrm{s}$. The fluctuations between this two water species could explain the thermodynamic and dynamic anomalies.  
Unfortunately, the most evident critical phenomena are expected to occur in the supercooled phase at temperature/pressure values not directly accessible in bulk water samples, and this prevents a direct experimental solution of the problem.  
In fact, in several experimental and simulation works the water singularity temperature $T_\mathrm{s}$ at atmospheric pressure has been extrapolated to be around 223-228 K, while homogeneous nucleation fixes the lowest reachable temperature at about 231 K for liquid water. Actually, in a macroscopic bulk sample the crystallization phenomena and several experimental difficulties limit the minimum temperature around 243 K. Nano-confinement has the advantage of preventing water freezing and of extending the experimental investigation of the supercooled phase down to very low temperatures; it is however still unclear to what extent the interactions with the pore surfaces modify the water properties~\cite{taschin_10,cucini_10,gallo_10,mancinelli_10,taschin_13b}. 
The hydrogen bond network largely determines the intra- and inter-molecular vibrational spectrum of supercooled water; thus the experimental investigation of the spectral features gives precious information on the local intermolecular structures and on their variation with the temperature/pressure conditions. The investigation of high frequency spectrum, 1000-4000 cm$^{-1}$, gives access to the intra-molecular vibrational dynamics that has been recently related to the LLCP hypothesis~\cite{mallamace_07}. The low frequency spectrum of water, $\nu < $ 1000 cm$^{-1}$, reflects the inter-molecular dynamics and so it is particularly sensitive to the local molecular structures. This frequency range of the water spectrum has been investigated in several light scattering studies, see for example \cite{krishna_83,aliotta_86,walrafen_86,rousset_90,mizoguchi_92,sokolov_95} and references at http://www.lsbu.ac.uk/water/. 
Two main broad peaks are observed at 50 and 200~cm$^{-1}$, generally attributed to the hydrogen bond ``stretching'' and ``bending'' vibrations. The spectra show also a very low frequency wing, $\nu <$ 20~cm$^{-1}$, that has been attributed to ``relaxation'' processes. The light scattering data were measured at relatively high temperatures and have rather poor signal-to-noise ratio, due to the very weak water signal: these drawbacks did not permit to identify all the possible vibrational modes possibly contributing to the water spectrum in this region. 
Moreover, the reported attempts of fitting the light scattering data have been based on a simple superposition of independent Gaussian and Lorentzian functions, neglecting any possible coupling between the vibrational dynamics and the structural relaxation processes. 

An accurate analysis of the low frequency spectral features of supercooled water, including a quantitative comparison to the predictions of recent models for water dynamics is still missing; it actually appears as a required step towards the full understanding of the structural and dynamical properties of water in its low temperature metastable state.

In recent years, time-resolved non-linear spectroscopy demonstrated to be a very useful tool for the investigation of complex liquid dynamics~\cite{torre_08}. 
Several studies of this type have been performed on liquid water; only in few cases have they been extended to its supercooled phase~\cite{torre_04,taschin_06,bartolini_09,taschin_11}. 
Time-resolved optical Kerr effect (OKE) investigations of liquid water~\cite{castner_95,palese_96,winkler_02,torre_04,ratajska_06} show the signature of fast vibrational dynamics at short times, followed by a slower monotonic relaxation. The initial oscillatory component provides information about intermolecular hydrogen-bond dynamics, and can be directly compared with the low-frequency Raman spectrum~\cite{castner_95}. Actually, the OKE experiments on supercooled water~\cite{torre_04} have been focused so far on the investigation of the slower decay, which is interpreted as a structural relaxation phenomenon. No detailed OKE investigation of the vibrational dynamics in supercooled water has been carried out, due to the intrinsic experimental difficulties, so that a detailed  picture of this dynamics is still missing~\cite{sonoda_05}.
Using an improved experimental set-up~\cite{bartolini_09}, we measured, for the first time according to our knowledge, the fast vibrational dynamics and the slow structural relaxation in a single data set characterized by low noise and wide dynamic range. Moreover, we reached temperatures lower than in previous experiments, approaching the homogeneous nucleation limit. We succeeded in measuring the OKE signal relaxation over a large time interval of  about 25 ps with a high time resolution, better than 20 fs. We can thus provide the unambiguous measurement of the entire correlation function in supercooled water, enabling a very stringent comparison with the available theoretical models.

\section*{Results}

\subsection*{Experimental techniques}

Time-resolved OKE spectroscopy is a non-linear technique exploiting the transient birefringence induced by the electric field of radiation in an optically transparent and isotropic medium. The method has been described in many papers and review articles~\cite{righini_93,hunt_07,bartolini_08}. In short, the oscillating electric field of a polarized short laser pulse (the pump) induces optical birefringence in the sample. In addition to the instantaneous electronic contribution (which does not carry any dynamic information), the induced birefringence reflects the relaxation and vibrational response of the molecules in the sample. A second pulse (the probe), spatially superimposed within the sample with the pump pulse and characterized by a different polarization, probes the induced birefringence. The intensity of probe light leaking through a polarizer placed after the sample measures the material response as a function of the time delay between pump and probe.

If femtosecond laser pulses are employed, the method enables to measure the relaxation processes and the low frequency vibrational correlation functions of the system. In this sense it is the time-domain counterpart of depolarized light scattering technique~\cite{hunt_07,bartolini_08}. The wide accessible time window, extending from tens of femtoseconds to hundreds of nanoseconds, makes OKE a very flexible technique, capable of revealing very different dynamic regimes, from simple liquids to supercooled liquids and glass formers~\cite{bartolini_99,torre_98,torre_00,prevosto_02,ricci_02,ricci_04}. 
In the case of optical heterodyne detection, the delay time dependent signal is directly proportional to the material response (relaxation) function, $R(t)$, convoluted with the instrumental function, $G(t)$, see Methods. The response $R(t)$ is directly connected to the time derivative of the time-dependent correlation function of the anisotropic component of the collective susceptibility\cite{fecko_02,prevosto_02,hunt_07,bartolini_08}: 
\begin{equation}
R(t)\propto \frac{\partial}{\partial t}\langle\chi(t)\chi(0)\rangle
\label{signaleq}
\end{equation}
The imaginary part of the Fourier transform of $R(t)$ corresponds to the frequency-dependent response, the observable measured in depolarized light scattering (DLS) experiments~\cite{kinoshita_95}. We stress here that measuring the correct instrumental function $G(t)$ is a crucial step for achieving a reliable and accurate fitting of the experimental OKE signal. A detailed description of the method adopted is given in the Methods section. 

Supercooling bulk water can be achieved only with particularly pure sample and using specific attention for the cooling procedure. We performed the measurements on a sealed vial of cylindrical shape, prepared for pharmaceutical purposes by the Angelini company, which allowed us to reach the temperature of 247 K. The sample temperature was controlled by means of a home-made cryostat with a stability of 0.1 K, using a Peltier cooler.
 
\subsection*{Measured data}

\begin{figure}
\begin{center}
\centerline{\includegraphics[scale=1.3]{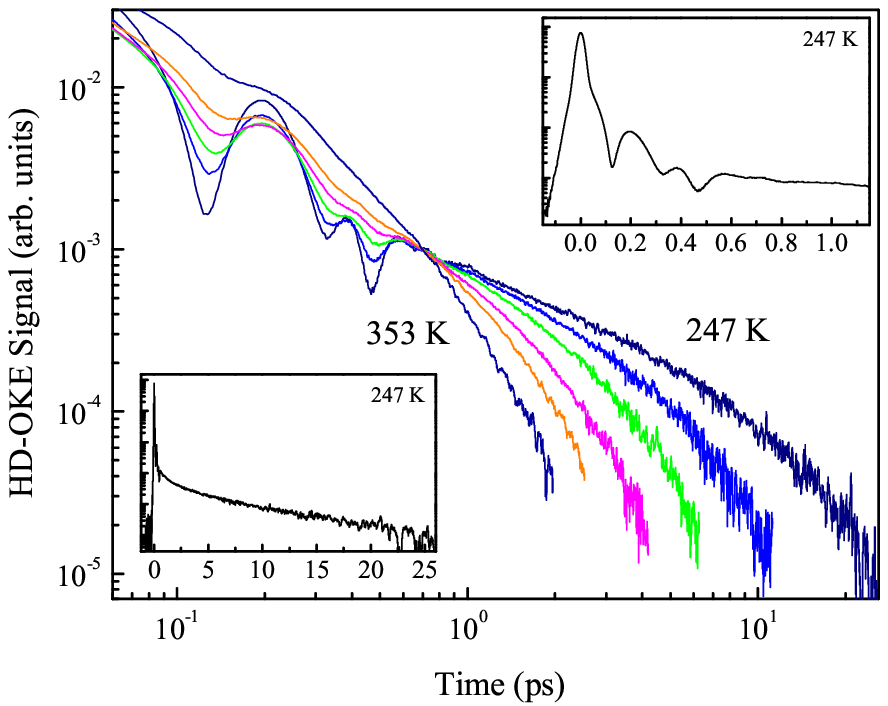}}
\caption{\textbf{HD-OKE signal from liquid and supercooled water} \\ In the main panel, the experimental signal is reported as function of temperature in a log-log plot. At short delay times ($<$1~ps), the heterodyne-detected optical Kerr effect (HD-OKE) signal shows oscillations due to vibrational dynamics, that become more defined as the temperature is lowered below the melting point. At longer delay times, the signal is characterised by a monotonous decay, which is strongly temperature dependent and clearly non-exponential. For a readable display, only 6 temperatures (247, 258, 273, 293, 313, 353~K) out of the 11 measured (the previous ones plus 253, 263, 268, 278, 333~K) are reported; the intensities are normalized at 0.7~ps. The upper and lower insets show, in a log-linear plot, the 247~K HD-OKE data at short and long delay times, respectively.}\label{fig_1}
\end{center}
\end{figure}

The heterodyne-detected OKE (HD-OKE) data are reported in Figure~\ref{fig_1}: the data clearly show the complex dynamic features presented by liquid and supercooled water. The vibrational dynamics, extending up to 1 psec, displays an articulate pattern that becomes progressively more defined going deep in the supercooled phase, and that at longer times merges into the monotonic decay due to structural relaxation.

The time-resolved HD-OKE experiment measures the time-dependent correlation function of the water susceptibility (i.e. collective polarizability), see Equation \eqref{signaleq}. The time evolution of this function reflects the water dynamics through the electronic polarizability. Numerical investigations~\cite{fecko_02,skaf_05,sonoda_05} show that the OKE signal of water is the result of a complex interplay between intrinsic molecular terms and interaction-induced contributions. These calculations reveal how the fluctuations of the water susceptibility, including the anisotropic components, are largely determined by the intermolecular translational dynamics. This behavior is expected, given the essentially isotropic water molecular  polarizability. The density fluctuations (i.e. translational dynamics) dominate the relatively low frequency range of the density-of-states\cite{padro_04,desantis_04}, $\nu < $ 400 cm$^{-1}$, whereas the rotational/librational dynamics affect the high frequency spectrum\cite{cho_94},  $\nu > $ 400 cm$^{-1}$. Moreover, the structure and dynamics of the H-bond network are fundamental properties that affect the susceptibility fluctuations at any frequency.
 
\begin{figure}
        \begin{center}
        \includegraphics[scale=0.9]{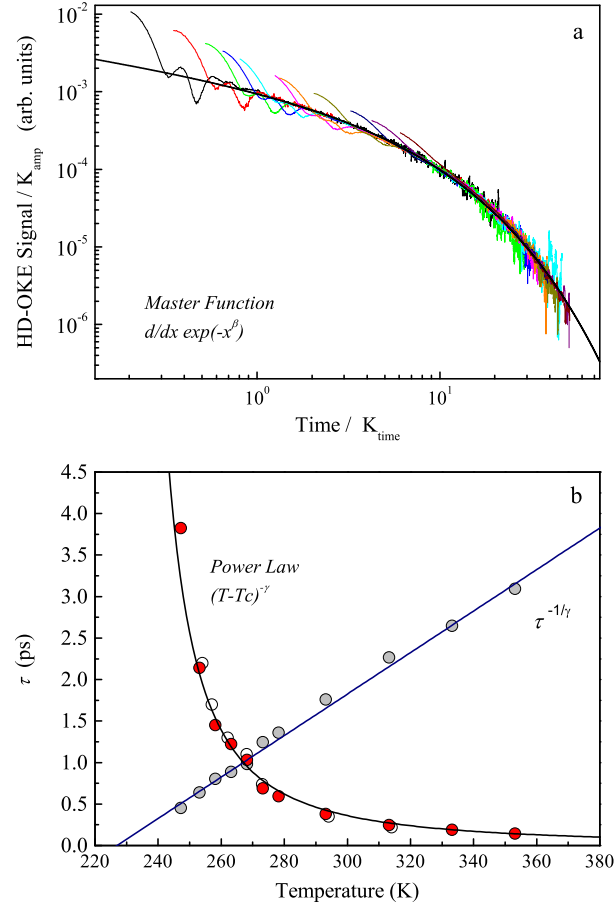}
              \caption{\textbf{Master-plot and scaling law} The slow dynamics of supercooled water can be analyzed using a model-independent procedure~\cite{torre_04}, consisting in rescaling the time/amplitude scales of the raw data measured at different temperatures in a way that the decay curves overlap. If all the data collapse in a ``master-plot'', it is possible to identify a function, the ``master-function'', which defines the relaxation independently of the temperature~\cite{goetze_92}. In the panel \textbf{a} we report the master-plot of the heterodyne-detected optical Kerr effect data and the master function: $\frac{d}{dx}exp(-x^\beta)$ with stretching parameter $\beta=0.6$. The relaxation times obtained with the rescaling procedure are shown in the panel \textbf{b} of the figure. The new data (red circles) extend and confirm the previous data (empty circles)~\cite{torre_04}. The temperature dependence of these data are in agreement with the power law: $\tau \propto (T-T_\mathrm{c})^{-\gamma}$, with critical temperature $T_\mathrm{c} \simeq 227$ K and critical exponent $\gamma \simeq 1.7$. We report the relaxation times and the power law in a linearized plot, in order to visualize the scaling proprieties and the critical temperature.}
        \label{fig_2}
        \end{center}
   \end{figure}

The relaxation features presented by supercooled water are very similar to those of glass-forming liquids~\cite{torre_00}, and the slow dynamics is found in agreement with the main mode-coupling theory predictions~\cite{torre_04}. In Figure~\ref{fig_2} we report the master plot analysis of the slow component of the HD-OKE signal decay. The new data confirm, over an extended temperature range, the power law temperature dependence of the relaxation time. Other experimental investigations reported similar power law temperature dependence of different water observables \cite{debenedetti_03b}, both static and dynamic; recently a small angle x-ray scattering investigation \cite{huang_10} found that also the correlation length of the density fluctuations follow a power law with similar value of $T_\mathrm{c}$.   

\subsection*{Data Analysis and Fits}

The analysis of the whole time-dependent correlation function measured by the optical experiments requires the utilization of a model that enables an operative parametrization of the vibrational modes present in the water dynamics. This turns out to be a complex task, and different attempts have been pursued~\cite{palese_96,winkler_02,ratajska_06} in that direction. None of these water studies was extended into the supercooled state. Moreover, the data analysis and the fitting models adopted so far did not include any coupling between the fast vibrational dynamics and the slow structural relaxation. We tested some of these fitting procedures on our HD-OKE data and found that they fail to reproduce correctly the HD-OKE data in the supercooled temperature range. In order to analyze the vibrational dynamics and the structural relaxation including their coupling processes, we utilized the schematic solution of mode-coupling theory \cite{goetze_92}. 

The formalism of mode-coupling theory (MC) stems from a generalization of the Mori and Zwanzig hydrodynamic theory \cite{goetze_92}. In MC theory the retardation effects are taken into account by memory functions $K(t)$, which play the role of the hydrodynamic friction coefficients. Of course the task of any microscopic theory is that of providing the form of $K(t)$. In its simplest form, developed for simple atomic liquids, only density fluctuations contribute to memory; in order to overcome the limitations of this approach and to extend the MC theory description to the dynamics of more complex systems, several versions of the theory have been developed. For molecular liquids in fact orientational and intra-molecular degrees of freedom contribute to the overall dynamics. In our analysis we adopt the multi-component schematic formulation of the MC theory (SMC)\cite{goetze_00b,goetze_04,goetze_09}. Briefly, in this approach the density-density time correlator, $\Phi_\mathrm{m}\propto\langle\rho(t)\rho(0)\rangle$, termed the master correlator, is described by the differential equation \cite{goetze_00b}: 
\begin{equation}
\ddot{\Phi}_\mathrm{m}(t)+\eta_\mathrm{m}\dot{\Phi}_\mathrm{m}(t)+\Omega_\mathrm{m}^2\Phi_\mathrm{m}(t)+
	\int_0^t K(t-t')\dot{\Phi}_\mathrm{m}(t')dt'=0,
\label{mct-eq}
\end{equation}	
	 where $\Omega_\mathrm{m}$ is the mode frequency, $\eta_\mathrm{m}$ is the friction coefficient and the memory function $K(t)= v_1 \Phi_\mathrm{m}(t)+v_2 \Phi^2_\mathrm{m}(t)$ is expressed as a series expansion of the master correlator itself. The role of other observables, $A_\mathrm{i}$, linked to the density correlator dynamics (e.g. the molecular orientation, the dynamics of a target molecule, etc.) is described by a set of differential equations of the same form as equation \eqref{mct-eq}, written in terms of slave correlators $\Phi_\mathrm{i}(t)\propto\langle A_\mathrm{i}(t)A_\mathrm{i}(0)\rangle$ and of memory functions of the form  $m_\mathrm{i}(t)= v_\mathrm{i}^\mathrm{s}\Phi_\mathrm{m}(t)\Phi_\mathrm{i}(t)$, which implies the coupling of the master and slave correlators.
The key point in the SMC analysis of the OKE experimental data is that the time correlation function of the sample polarizability can be expressed as the sum of the time derivatives of the slave correlators:
\begin{equation}
R^\mathrm{SMC}(t)\propto\frac{\partial}{\partial t}\sum_\mathrm{i}\Phi_\mathrm{i}(t).
\label{mct-response}
\end{equation}
We underline here that the individual slave correlators describe ``average collective modes'', so that they cannot be associated to specific physical variables.
In summary, the SMC approach represents a robust physical model capable of describing a complex dynamics, with no need of unjustified assumptions concerning the decoupling of slow and fast dynamics. 

We solved the SMC equations numerically, taking the frequencies, friction and coupling coefficients as parameters to be adjusted in order to reproduce the HD-OKE response by a fitting procedure. The details on the SMC equations and fitting procedures are reported in the Methods section. 

\begin{figure}
        \begin{center}
       	\includegraphics[scale=0.9]{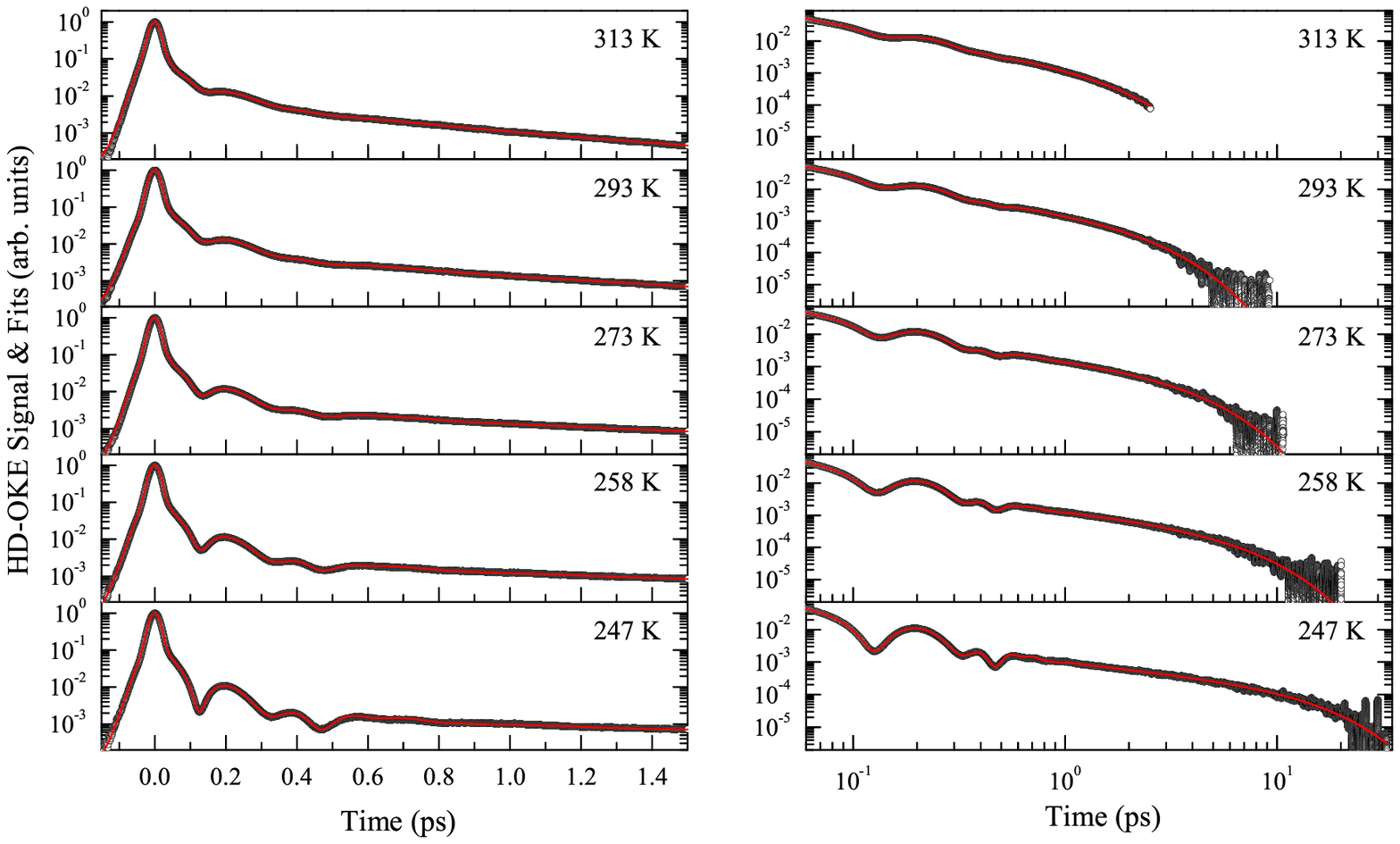}
        \caption{\textbf{Data analysis and fit based on SMC theory} \\The complex vibrational dynamics taking place in the sub-picosecond time scale is correctly reproduced by the numerical solution of the schematic mode-coupling (SMC) model. The SMC model enables a very good fit of the slow relaxation decays, corresponding to the stretched exponential function. Also at intermediate delay times, where the vibrational dynamics merge into the relaxation processes, the decay curve is correctly described by the SMC equations. We report the SMC fits (red line) of the heterodyne-detected optical Kerr effect data (circle) for five temperatures. In the left panel the short time window is shown in a log-linear plot, in the right panel the whole time window is shown in a log-log plot.}         
        \label{fig_3}
       \end{center}
\end{figure}

In Figure~\ref{fig_3} we show some of the measured HD-OKE signals with the fits obtained using the SMC model. The model reproduces correctly the experimental data over the whole time range at all temperatures. Most remarkably, and differently from other fitting models widely used, this result is achieved without imposing any non-physical decoupling of vibrational and relaxation dynamics. 

An important result of our work is that the fitting at temperatures $\leq$ 313 K requires three slave correlators, whereas for higher temperatures only two correlators are needed. We would like to stress out that the master correlator alone cannot reproduce the data suggesting that the density fluctuations are not enough to describe completely the dynamics of collective water polarizability; as expected the hydrogen bound dynamics adds not trivial dynamical components that determine the water susceptibility fluctuations.     

\begin{figure} 
       \begin{center}
      	\includegraphics[scale=0.9]{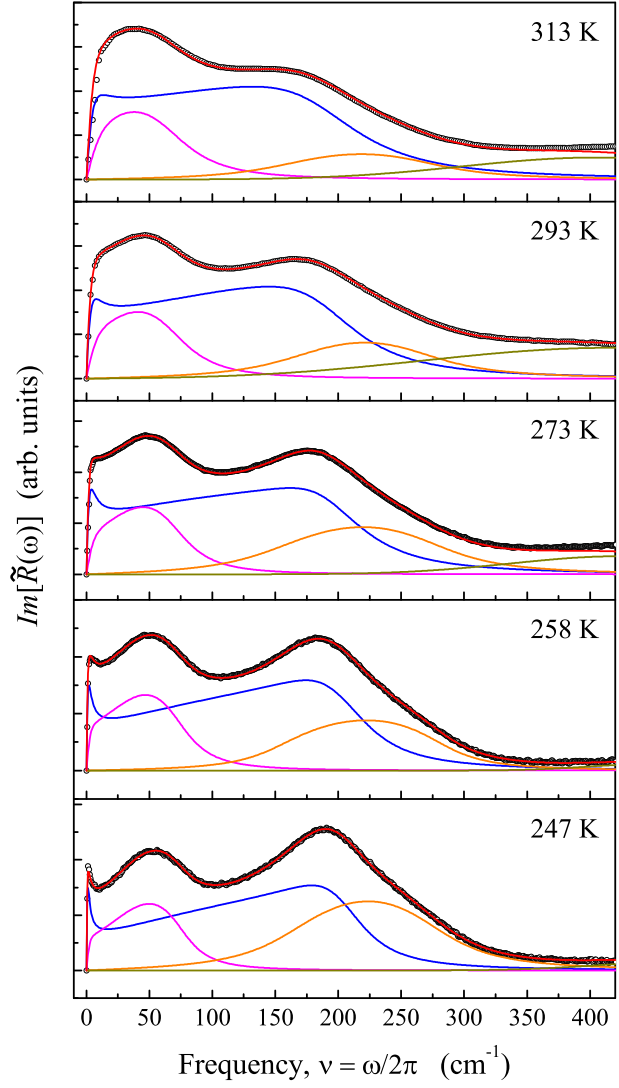}
        \caption{\textbf{Water spectrum and vibrational components} The Fourier transform of the time-dependent optical Kerr effect (OKE) response, $\tilde{R}(\omega)$, is directly related to the frequency-dependent spectrum measured in the dynamic light scattering  experiments. The imaginary part of the Fourier transform of the heterodyne-detected OKE data (open circle), after that the contribution of the instrumental function has been removed by de-convolution, and of schematic mode-coupling (SMC) fits (red line) at 247, 258, 273, 293, 313~K are reported. The SMC model enables to disentangle the dynamical features that generate the simulated response function. The contributions of the three slave correlators are reported in the figure as magenta-blu-orange lines. Moreover, a simple damped harmonic oscillator (dark yellow) has been added to the SMC response to reproduce the very high frequency vibrational contributions ($\nu >$ 400 cm$^{-1}$); this does not affect the response function in the lower frequency range, where the dynamics relevant to our investigation takes place. The simulation of HD-OKE data by SMC model requires two vibrational modes in order to fit the intermolecular stretching band of water (blu and orange lines). The characteristics of these two modes are clearly different in terms of spectrum shape and temperature dependence.} 
        \label{fig_4}
      \end{center}
\end{figure}

The time domain data, collected in the time-resolved HD-OKE experiments, are transformed to yield the corresponding spectra in the frequency domain. In fact, the Fourier transform of the HD-OKE response function, $\tilde{R}(\omega)$, is directly connected with the spectrum $S_\mathrm{DLS}(\omega)$ measured in the depolarized light scattering (DLS) experiments: the imaginary part of $\tilde{R}(\omega)$ is proportional to $S_\mathrm{DLS}(\omega)$ corrected for the Bose factor, i.e $Im[\tilde{R}(\omega)] \propto S_\mathrm{DLS}(\omega)/[n(\omega)+1]$~\cite{kinoshita_95,hunt_07,bartolini_08}.
Our results then can be directly compared to the DLS, Rayleigh and Raman data, found in the literature. 
The imaginary part of the frequency domain response function, $Im[\tilde{R}(\omega)]$, was obtained as the Fourier transform of the HD-OKE signal deconvoluted from the instrumental function: $Im[\tilde{R}(\omega)] \propto Im\lbrace FT\left[ S(t) \right] /FT\left[ G(t)\right] \rbrace$. The Fourier transform of the measured HD-OKE response and of the calculated SMC response are reported in Figure~\ref{fig_4}.

\section*{Discussion}

The experimental spectra clearly show the characteristic vibrational features of liquid water: the two broad bands around 50 and 200~cm$^{-1}$, and the low frequency wing. This wing corresponds to the slow decay of the OKE signal measured in the time domain; as discussed in our previous work~\cite{torre_04}, it is attributed to structural relaxation.

The other two bands are generally attributed to translational intermolecular modes (i.e. the oscillatory dynamics of water molecules around their instantaneous center of mass) hindered by the first neighbour molecular cage~\cite{skaf_05,desantis_04,padro_04}. In the liquid phase they are largely affected by the hydrogen bonds. A comparison with the calculated vibrational density-of-states reported in ref.~\cite{desantis_04} suggests specific assignment of the 50 and the 200 cm$^{-1}$ bands. The first one is dominated by the transverse translational motion, while the second has essentially longitudinal character~\cite{desantis_04}. The physical interpretation is that the two intermolecular vibrations are mainly involving ``bending'' and ``stretching'' distortion of the hydrogen bonds, respectively~\cite{walrafen_86}. This picture even if intuitively useful, is clearly oversimplified:  it doesn't take into account the coupling of the two vibrations imposed by the complex hydrogen bond network present in the liquid water. The presence of two different local structures, the one corresponding to highly tetrahedral intermolecular structures, and that consisting of strongly distorted network, as well as their continuous reorganization induced by the structural relaxation processes, are expected to largely affect the vibrational dynamics.  

The SMC fit enables to disentangle the vibrational modes contributing to this dynamics and to point out their coupling to the structural relaxation processes. The stretching mode, around 200 cm$^{-1}$, needs two correlators to be properly reproduced. One correlator, peaked at about 180 cm$^{-1}$ (blue line), shows an almost temperature independent peak amplitude and a large spectral component extending to very low frequency, indicative of strong coupling to the structural dynamics. The other correlator, peaked at about 225 cm$^{-1}$ (orange line), appears to be substantially uncoupled to the structural dynamics; its amplitude increases with decreasing temperature. Our HD-OKE measurement represents the first clear experimental determination of this feature, even if it was inferred in a previous Raman study\cite{krishna_83}. The bending mode at about 50 cm$^{-1}$ can be described by a single correlator (magenta line in Figure~\ref{fig_4}) that turns out to be decoupled from the structural relaxation (the peak close to 0 frequency). It is possible that also the bending peak develops at low temperature a second component, but very likely the frequency separation is not sufficient to be resolved from our data analysis.

A comparison between these results and the recent debate on the nature of water anomalies is due. Our data and analysis show the presence of two distinct modes responsible for the part of the water spectrum normally attributed to the inter-molecular stretching vibrations. Raman spectra of low and high-density amorphous ice (LDA and HDA) in this frequency region differ from that of liquid water~\cite{suzuki_00}. LDA has a broad peak centred at higher frequency, about 220 cm$^{-1}$, HDA shows an even broader peak centered at about 185 cm$^{-1}$ (see for example figure 1 in ref.~\cite{suzuki_00}) whereas the crystalline ice phases typically have a narrow peak at 230 cm$^{-1}$. This suggests that the modes observed in our experimental investigations can be associated with two fluctuating water species with different local structures; a low-density form (LD) characterized by tetrahedral network and four-coordinated molecules similar to the LDA structure, and a high-density (HD) form characterized by closely packed aggregates with lower coordination and high network distortions. The mode at 225 cm$^{-1}$ can be attributed to the LD form and the mode at 180 cm$^{-1}$ to the HD one. The present data strongly support the existence of the LD and HD water species. Nevertheless, as the observed water dynamic is measured at temperature/pressure relatively far from the supposed LLCP, some ambiguity remains about the nature of the critical phenomena taking place into the supercooled phase. The spectroscopic features attributed to the LD and HD forms could be attributed either to the coexistence of two extended liquid phases leading to the liquid-liquid phase transition\cite{poole_92,mishima_98,debenedetti_03b}, or to the continuous transition from a normal, unstructured liquid to a tetrahedrally structured liquid\cite{moore_09}.

Our new experimental data and analysis clearly show the presence of the high-frequency satellite mode contributing to the intermolecular stretching band of water, whose amplitude increases noticeably when the sample is cooled into the supercooled metastable phase. The nature of this mode, peaked at 225 cm$^{-1}$, appears clearly different from that of the nearby mode at 180 cm$^{-1}$. In fact, the former is strongly temperature dependent and only weakly coupled to the structural relaxation processes, while the latter is almost temperature independent and strongly coupled to structural rearrangements. The surprising similarities of the spectra of LDA and HDA glass forms with those of the two high frequency distinct modes, reported in this work, suggest attributing them to the LD and HD water structures, respectively. The LD water forms are expected to consist of aggregates of a small number of molecules; being mostly localized in space (at least in the temperature range investigated), their dynamics show negligible coupling with the structural relaxation. The two local configurations of liquid water coexist in the entire temperature interval considered, above and below the melting temperature. The number/lifetime of the LD species must increase with decreasing temperature in order to account for the amplitude increase of the corresponding correlator revealed by our analysis. On the other hand, the vibrational dynamics of the fraction of HD water appears dominated by the bare caging effect that merges into the structural relaxation, a behaviour typical of fragile glass-former liquids.

We end by remarking that our findings are consistent with the conclusion drawn recently by Nilsson et al.\cite{nilsson_12} on the basis of x-ray emission and absorption experiments, and by Overduin and Patey\cite{overduin_12} from the analysis of their computer simulations. In both papers large density fluctuations in liquid water are attributed to a short range bimodal aggregation of molecules with disordered and tetrahedral local environments.

\section*{Methods}

\subsection*{Water Sample}

Supercooling bulk water is not an easy task because of crystallization. Temperatures as low as 20 degrees below melting can be achieved only with particularly pure sample and using specific attention for the cooling procedure. In our work we performed the measurements on bi-distilled water prepared by the Angelini company in a sealed vial of cylindrical shape, which enables us to reach a temperature as low as 247 K. The sample temperature is controlled with a stability of 0.1 K by means of a home-made cryostat. The vial is inserted into a parallelepiped-shaped aluminum holder, with a central cylindrical housing whose diameter is close to that of the vial. Two fused silica windows on opposite sides of the aluminum holder allow the beams to cross the sample. A thin film of glycerol between the vial and the housing assures an efficient heat transfer. The holder is fixed to the cold plate of a Peltier cooler and the temperature is controlled by a platinum thermoresistance in thermal contact with the holder itself.

\subsection*{HD-OKE set-up}

\begin{figure}
\begin{center}
\includegraphics[scale=0.8]{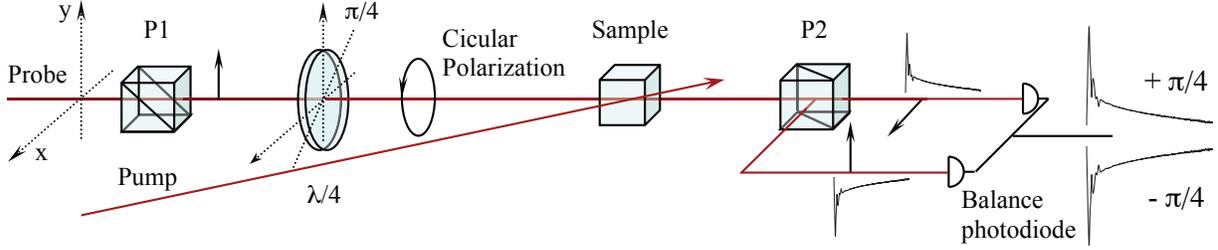}
\caption{\textbf{The optical set-up for the heterodyne-detected optical Kerr effect measurements (HD-OKE)}. We adopted the optical configuration with circularly polarized probe beam and differential acquisition of two opposite-phase signals on a balanced double photodiode. A quarter wave-plate placed between the two polarizers produces a circularly polarized probe field. Two signals, corresponding to the horizontal and vertical linear polarizations generated by polarizer P2 and having a opposite sign, are sent to the balanced photodiode detector. Thus the measured OKE signal is automatically heterodyne-detected and free from possible spurious signals.
The delay time between the pump and probe laser pulses was scanned by means of the continuous motion of a translation stage, whose absolute position was controlled by a linear encoder. This method reduces the acquisition time substantially, while improving the signal statistics. It enables to measure the fast vibrational dynamics and the slow structural relaxation in a single experiment, with high signal-to-noise ratio.
A further improvement of the data quality was obtained by subtracting the two measurements, corresponding to left and right circular polarizations of the probe field.}
\label{fig_5}
\end{center}       
\end{figure}

We used the optical configuration introduced by Giraud et al.\cite{giraud_03}, which enables a reliable and efficient heterodyne-detection of the HD-OKE signal; a scheme of the optical set-up is shown in Figure \ref{fig_5}. The laser system utilized is a self-mode-locked Ti:Sapphire laser (by Femtolasers, mod. Fusion) producing pulses of 20 fs duration with an energy of 3 nJ; 77.47 MHz repetition rate, 800 nm wavelength. 
Other details on the experimental set-up are reported in ref.~\cite{bartolini_09,bartolini_07}.

\subsection*{HD-OKE signal}

The signal measured in an HD-OKE experiment is~\cite{hellwarth_77,mcmorrow_88,torre_93}:
\begin{equation}
	S(t)\propto\int_{-\infty}^{+\infty} \left[ \gamma\delta(t-t')+R(t-t') \right]  G(t') dt'
\label{signalfit}\
\end{equation}
where $\gamma\delta(t)$ is the instantaneous electronic response and and $R(t)$ is the nuclear response, see eq.\ref{signaleq}. $G(t)$ is the instrumental function accounting for the finite time resolution determined by the time width of the pump and probe pulses. Under specific conditions~\cite{mcmorrow_88}, $G(t)$ corresponds to the temporal cross-correlation of the laser pulse intensities. 
 
\subsection*{Instrumental function measurements}

The measurement of the real instrumental function $G(t)$ is mandatory to extract the true OKE response of water but it is far from trivial. By definition, the instrumental function in a HD-OKE experiment is the signal measured in a system with response function $R(t)\propto\delta(t)$, as directly inferred from eq.\ref{signalfit}. In practice, this condition can be fulfilled by choosing a sample with negligible or instantaneous nuclear response. In most of the papers published on this subject in recent years a fused silica plate has been used to this purpose, as its nuclear contribution is considered negligible. In reality, silica presents a fast and weak, but not null, nuclear contribution that prevents from measuring the real instrumental function. Alternatively, the instrumental function has been obtained by measuring the second harmonic cross-correlation of the laser pulses\cite{kinoshita_95}. This procedure, however, implicates some changes and adjustments in the OKE set-up that potentially modify the response function. Actually, any even small changes of the experimental conditions (e.g. a different beam alignment or a modified angle of superposition of the beams inside the sample) affect the instrumental function. Therefore, any method for obtaining the instrumental function which involves changes of the set-up is unsafe and can yield an incorrect time profile. 
In other HD-OKE investigations the response function has been obtained simply by fitting the electronic contribution to the HD-OKE signal with an analytic peak function (like Gaussian, hyperbolic secant, etc.). 
None of these methods enable the measurement of response function with the accuracy required by HD-OKE investigations in the water sample, which is characterized by a fast and low level signal.
 
We extracted the instrumental function by performing HD-OKE measurements in a reference sample. This procedure allows us to accurately preserve the experimental conditions used during the water measurements. As the reference sample we chose a plate of calcium fluoride (CaF$_2$). This a cubic ionic crystal which is optically isotropic and, in the frequency range probed in our experiment, has only one Raman active band at 322~cm$^{-1}$, due to the optical phonon T$_\mathrm{2g}$ symmetry. The calcium fluoride plate was dipped inside a water vial, identical to that used for water measurements, supported by the same sample holder.

The measurements of the instrumental function were done by just replacing the water vial with the water-CaF$_2$ vial, leaving the rest of the set-up unchanged. The thickness of the CaF$_2$ plate, 3 mm, was enough to fully contain the probe and pump overlap volume, avoiding any spurious signal contribution by the outer water. We took a reference measurement for each set of water data. Maximum attention was taken to preserve the experimental conditions between the water and the reference measurement. 

\begin{figure}
\begin{center}
\includegraphics[scale=0.7]{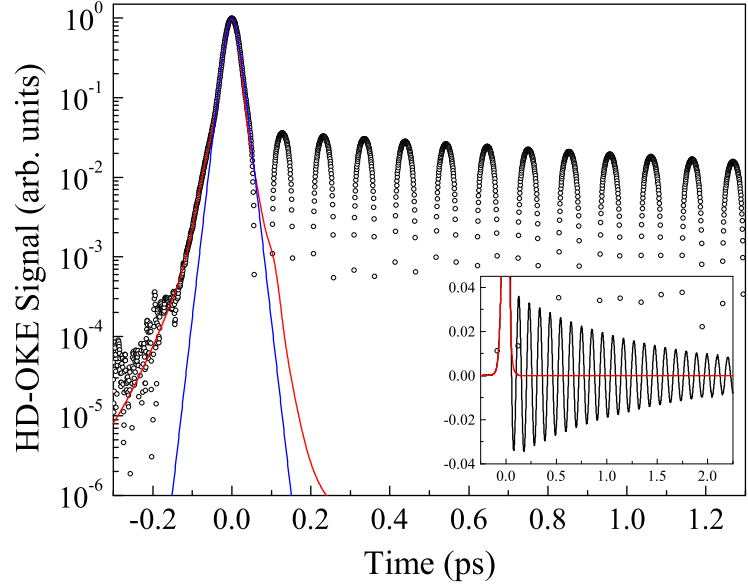}
\caption{\textbf{A typical HD-OKE signal of the reference sample and the extracted instrumental function}. A critical experimental point is that any minor, apparently negligible adjustment of the optical set-up, can hampers the extraction of the real instrumental function, $G(t)$, and hence the accurate deconvolution of the HD-OKE signal. In our experiments $G(t)$ was obtained by measuring the HD-OKE signal of a CaF$_2$ plate placed in a purposely designed holder which enables switching to the water sample without even touching the rest of optical set-up. In the figure we show the HD-OKE signal of CaF$_2$ (circles) and the instrumental function (red line) obtained with the fitting procedure described in the Methods. The nuclear response, $R$, is modelled as the time-derivative of a single damped oscillator; and the instrumental function, $G$, is simulated as the sum of Gaussian, Lorentzian, and hyperbolic secant functions. The comparison with the fit of the instantaneous electronic part performed using just a single hyperbolic secant (blue line) is reported too.}
\label{fig_6}
\end{center}
\end{figure}

\begin{figure}
\begin{center}
\includegraphics[scale=0.7]{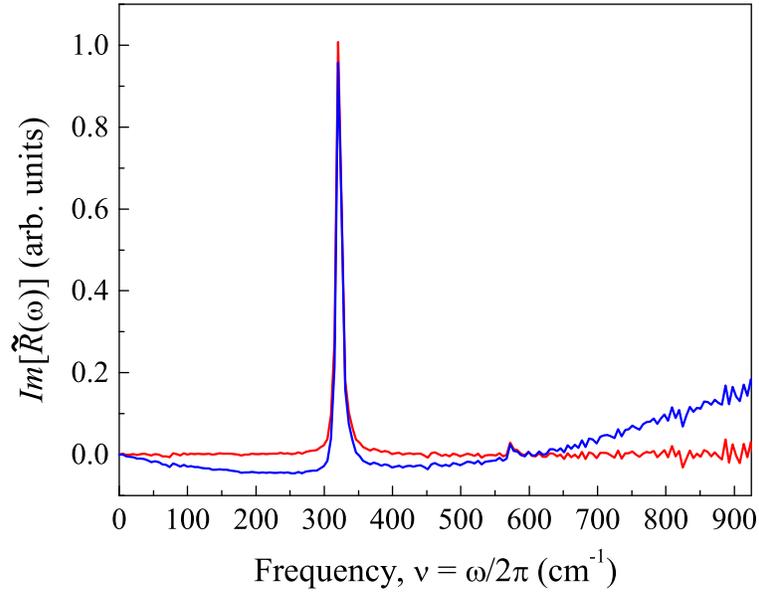}
\caption{\textbf{Dependence of the reference sample spectrum from the instrumental function choice}. The frequency domain response function $\tilde{R}(\omega)$ was obtained as the Fourier transform of the HD-OKE signal deconvoluted from the instrumental function: $Im[\tilde{R}(\omega)] \propto Im\lbrace FT\left[ S(t) \right] /FT\left[ G(t)\right] \rbrace$. As reference sample we used a calcium fluoride, CaF$_2$, crystal plate whose spectrum is well known and characterized by a single intermolecular vibration. The simplicity of the CaF$_2$ response enables a reliable test of the instrumental function extracted from the time domain data. The figure compares the CaF$_2$ frequency dependent response calculated using the instrumental function obtained by the iterative fitting procedure (red line) to the one obtained by fitting the electronic peak with a hyperbolic secant function (blue line). Only the knowledge of the real instrumental function allows measuring the correct response.}
\label{fig_7}
\end{center}
\end{figure}

Figure \ref{fig_6} reports a typical HD-OKE signal obtained in the CaF$_2$ reference sample. We performed a least square fitting of the HD-OKE data of the reference sample with the simulated signal, according to eq.\ref{signalfit}, where the nuclear response is a single harmonic damped oscillator. We found that the instrumental function requires the sum of several analytical functions, typically a combination of Gaussian, Lorentzian and/or hyperbolic secant functions. In Figure \ref{fig_6} we report the instrumental function obtained by the iterative fitting procedure (continuous line) and the one obtained by fitting the electronic peak with a simple hyperbolic secant function (dashed line). Apparently these two instrumental functions are very similar however the small differences cannot be neglected for an accurate investigation of water fast HD-OKE response. This is mostly evident when the data are Fourier transformed to the frequency domain. In fact obtaining the imaginary part of the frequency dependent response from the time domain data involves the division by the Fourier transform of the instrumental function. 
Figure \ref{fig_7} shows the comparison between the response calculated using the instrumental function obtained by the iterative fitting procedure (continuous line) and the one obtained by simply fitting the electronic peak with a hyperbolic secant function (dashed line). We clearly see that only an accurate measurement of the real time domain instrumental function yields the correct response. 

To our knowledge, this aspect of the OKE data analysis has never been addressed in such a quantitative way, and the present level of accuracy in analyzing the role of the instrumental function in HD-OKE experiments is unprecedented in recent literature. In particular, we want to stress that the knowledge of the correct instrumental function is fundamental in order to extract the true OKE response, especially for the short delay time range. This is particularly critical for weak signals characterized by complex relaxation dynamics, as in the case of water.

\subsection*{Data Analysis and Fits}

According to eq.\ref{mct-response}, in our analysis of the HD-OKE data we are decomposing the electronic polarizability correlator, $\Phi_{\chi\chi}$, into the sum of $\Phi_\mathrm{i}(t)$ correlators. It is implicit that the individual slave correlators cannot be directly associated to specific physical variables. Each of these correlators in fact describes an ``average collective mode" whose dynamics is described by the SMC equations. The vibrational and relaxation features and coupling processes are included in the SMC equations by definition. In this respect the present SMC equations represent a robust physical model capable of describing a complex dynamics including vibrational and structural relaxations, implicitly accounting for their mutual coupling. Differently from other approaches, the method does not require any decoupling or dynamic separation between the fast (vibrational) and the slow (relaxation) processes. Thus, once the temporal relaxation function of the master correlator is known, it can be used to calculate the relaxation of the slave correlators and, eventually, that of the OKE signal.

\begin{figure}
\begin{center}
\includegraphics[scale=0.7]{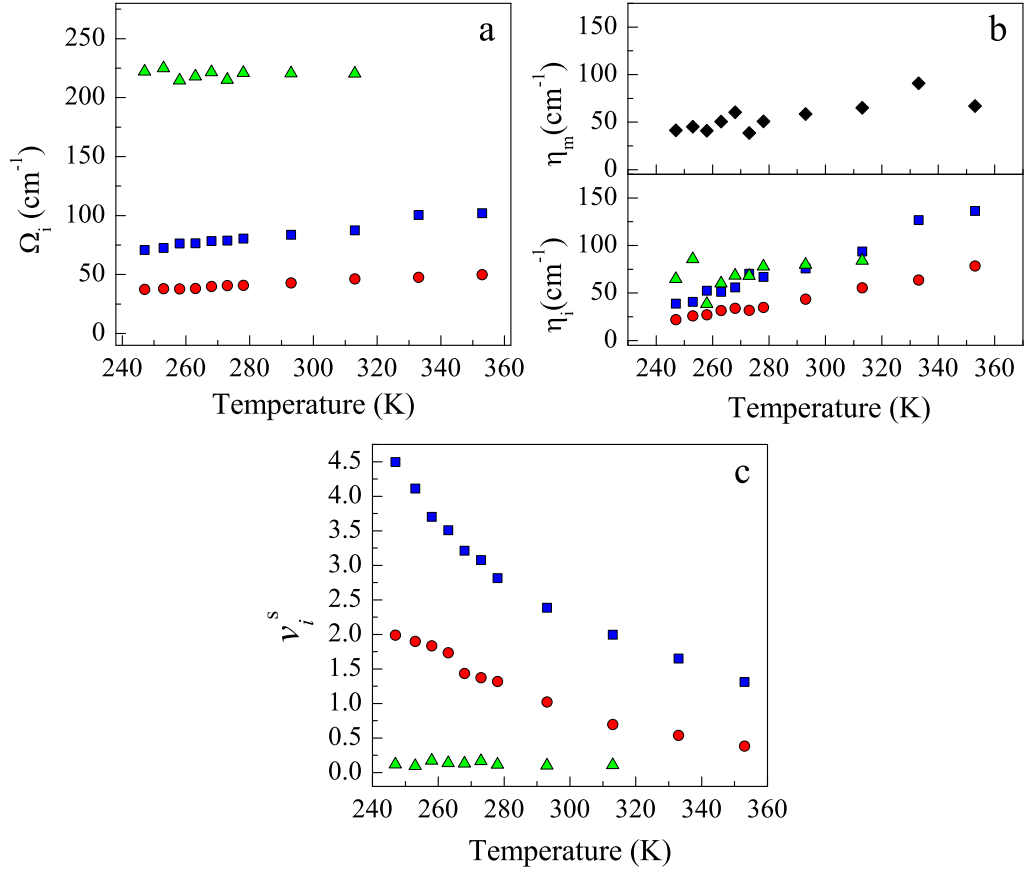}
\caption{\textbf{The temperature dependence of SMC fitting parameters}. The multi-component SMC model represents a robust physical model capable of describing a complex dynamics including vibrational and structural relaxations, implicitly accounting for their mutual coupling. We solved the SMC equations numerically, taking the frequencies, friction and coupling coefficients as parameters to be adjusted in order to reproduce the HD-OKE response by a fitting procedure. In particular, we found that the frequency $\Omega_\mathrm{m}$ and the vertex $v_1$ of the master oscillator are almost constant in the whole temperature range, and we locked their values to 66~cm$^{-1}$ and $0.33$, respectively. The second vertex $v_2$, instead, was found to increase almost linearly with decreasing temperature, and it was forced to obey the linear dependence $v_2=6-0.014\times T$. The friction parameter $\eta_\mathrm{m}$ and the remaining parameters of the slave oscillators were left free of varying.
In the figure we report the best fit values of the parameters. Panel~\textbf{a} reports the frequencies of the slave oscillators (circles, squares, and triangles refer to $\mathrm{i}=1$, $2$, and $3$, respectively). In panel~\textbf{b} we show the temperature behavior of the master and slave friction parameters (diamond refers to the master oscillator, circles, squares, and triangles refer to $\mathrm{i}=1$, $2$, and $3$, respectively). The temperature dependence of the vertices $v^\mathrm{s}_\mathrm{i}$ is reported in panel~\textbf{c} (circles, squares, and triangles refer to $\mathrm{i}=1$, $2$, and $3$, respectively).}
\label{fig_8}
\end{center}
\end{figure}

We found that the OKE data can be described in the entire temperature range by employing three slave correlators at most. In particular, the weight of the highest frequency contribution decreases monotonically with increasing temperature, becoming negligible at the two highest temperatures, where only two slave correlators are sufficient to reproduce the data. The parameters of the model are sixteen: the master equation parameters $\eta_\mathrm{m}$, $\Omega_\mathrm{m}$, $v_1$ and $v_2$, the slave equations parameters $\eta_\mathrm{i}$, $\Omega_\mathrm{i}$, $v_\mathrm{i}^\mathrm{s}$ with $\mathrm{i}=1,2,3$ and finally the three amplitudes $a_1$, $a_2$, and $a_3$. The integro-differential equations were solved numerically with a step by step second order Runge-Kutta algorithm.
Preliminary series of fits provided a qualitative estimate of the temperature behavior of the parameters. In agreement with similar analyses reported in the literature \cite{alba_95,krakoviack_02,wiebel_02,goetze_04}, in the subsequent step we forced some of the master parameters either to assume fixed values or to follow simple temperature trends, leaving all the slave parameters free.  
In Figure \ref{fig_8} we show the temperature dependence of the slave frequencies $\Omega_\mathrm{i}$ and the friction parameters $\eta_\mathrm{m}$ and $\eta_\mathrm{i}$, and of the three vertices $v^\mathrm{s}_\mathrm{i}$. 
The temperature behavior of the fitting parameters is consistent with the results obtained by the previous studies in glass-forming liquids~\cite{alba_95,krakoviack_02,wiebel_02,goetze_04}.

\subsection*{Author Contributions}
The contribution from all the authors was fundamental to realize this study. More specifically: A.T., P.B., R.E. and R.T. planned the experiments; A.T., P.B., R.E. performed the experiments and analysed data; R.R. and R.T. interpreted the data and write the paper in collaboration with the other authors.

\begin{acknowledgments}
The research has been performed at LENS. 
This work was supported by REGIONE TOSCANA POR-CRO-FSE 2007-2013 by EC COST Action MP0902-COINAPO and by PRIN 2010/2011 N.2010ERFKXL.
We would like to thank A. Marcelli for helpful discussion. We acknowledge M. De Pas, A. Montori and M. Giuntini for providing their continuous assistance in the set-up of the electronics. R. Ballerini and A. Hajeb for the accurate mechanical realizations.
\end{acknowledgments}



\begin{thebibliography}{10}

\bibitem{mishima_98}
O.~Mishima and H.E. Stanley.
\newblock The relationship between liquid, supercooled and glassy water.
\newblock {\em Nature}, 396:329--335, 1998.

\bibitem{debenedetti_03}
P.G. Debenedetti and H.~Eugene Stanley.
\newblock Supercooled and glassy water.
\newblock {\em Physics Today}, 56:40--46, 2003.

\bibitem{debenedetti_03b}
P.G. Debenedetti.
\newblock Supercooled and glassy water.
\newblock {\em J. Phys.: Cond. Matt.}, 15:R1669--R1726, 2003.

\bibitem{soper_08}
A.K. Soper.
\newblock Structural transformations in amorphous ice and supercooled water and
  their relevance to the phase diagram of water.
\newblock {\em Mol. Phys.}, 106:2053--2076, 2008.

\bibitem{angell_08}
C.~A. Angell.
\newblock Insights into phases of liquid water from study of its unusual
  glass-forming properties.
\newblock {\em Science}, 319:582--587, 2008.

\bibitem{nilsoon_11}
A.~Nilsoon and L.G.M. Pettersson.
\newblock Perspective on the structure of liquid water.
\newblock {\em Chem. Phys.}, 389:1--34, 2011.

\bibitem{moore_09}
Emily~B. Moore and Valeria Molinero.
\newblock Growing correlation length in supercooled water.
\newblock {\em J. Chem. Phys.}, 130:244505, 2009.

\bibitem{moore_11}
Emily~B. Moore and Valeria Molinero.
\newblock Structural transformation in supercooled water controls the
  crystallization rate of ice.
\newblock {\em Nature}, 479:506--508, 2011.

\bibitem{limmer_11}
David~T. Limmer and David Chandler.
\newblock The putative liquid-liquid transition is a liquid-solid transition in
  atomistic models of water.
\newblock {\em J. Chem. Phys.}, 135:134503, 2011.

\bibitem{sciortino_11}
F.~Sciortino, I.~Saika-Voivod, and P.~H. Poole.
\newblock Study of the st2 model of water close to the liquid-liquid critical
  point.
\newblock {\em Phys. Chem. Chem. Phys.}, 13:19759--19764, 2011.

\bibitem{liu_12}
Yang Liu, Jeremy~C. Palmer, Athanassios~Z. Panagiotopoulos, and Pablo~G.
  Debenedetti.
\newblock Liquid-liquid transition in st2 water.
\newblock {\em J. Chem. Phys.}, 137:214505, 2012.

\bibitem{kesselring_12}
T.~A. Kesselring, G.~Franzese, S.~V. Buldyrev, H.~J. Herrmann, and H.~E.
  Stanley.
\newblock Nanoscale dynamics of phase flipping in water near its hypothesized
  liquid-liquid critical point.
\newblock {\em Scientific reports}, 2:474, 2012.

\bibitem{gallo_12}
Paola Gallo and Francesco Sciortino.
\newblock Ising universality class for the liquid-liquid critical point of a
  one component fluid: A finite-size scaling test.
\newblock {\em Phys. Rev. Lett.}, 109:177801, 2012.

\bibitem{poole_13}
Peter~H. Poole, Richard~K. Bowles, Ivan Saika-Voivod, and Francesco Sciortino.
\newblock Free energy surface of st2 water near the liquid-liquid phase
  transition.
\newblock {\em J. Chem. Phys.}, 138:034505, 2013.

\bibitem{poole_92}
Peter~H. Poole, Francesco Sciortino, Ulrich Essmann, and H.~Eugene Stanley.
\newblock Phase behaviour of metastable water.
\newblock {\em Nature}, 360:324--328, 1992.

\bibitem{taschin_10}
A.~Taschin, R.~Cucini, P.~Bartolini, and R.~Torre.
\newblock Temperature of maximum density of water in hydrophilic confinement
  measured by transient grating spectroscopy.
\newblock {\em Europhys. Lett.}, 92:26005, 2010.

\bibitem{cucini_10}
R.~Cucini, A.~Taschin, P.~Bartolini, and R.~Torre.
\newblock Acoustic, thermal and flow processes in a water filled nanoporous
  glass by time-resolved optical spectroscopy.
\newblock {\em J. Mech. Phys. Solids}, 58:1302--1317, 2010.

\bibitem{gallo_10}
P.~Gallo, M.~Rovere, and S.-H. Chen.
\newblock Dynamic crossover in supercooled confined water: understanding bulk
  properties through confinement.
\newblock {\em J. Phys. Chem. Lett.}, 1:729, 2010.

\bibitem{mancinelli_10}
R.~Mancinelli, F.~Bruni, and M.A. Ricci.
\newblock Controversial evidence on the point of minimum density in deeply
  supercooled confined water.
\newblock {\em J. Phys. Chem. Lett.}, 1:1277, 2010.

\bibitem{taschin_13b}
Andrea Taschin, Paolo Bartolini, Agnese Marcelli, Roberto Righini, and Renato
  Torre.
\newblock A comparative study on bulk and nanoconfined water by time-resolved
  optical kerr effect spectroscopy.
\newblock {\em Faraday Discussions}, page DOI:10.1039/C3FD00060E, 2013.

\bibitem{mallamace_07}
Francesco Mallamace, Matteo Broccio, Carmelo Corsaro, AntonioFaraone, Domenico
  Majolino, Valentina Venuti, Li~Liu, Chung-Yuan Mou, and Sow-Hsin Chen.
\newblock Evidence of the existence of the low-density liquid phase in
  supercooled,confined water.
\newblock {\em Proc. Natl Acad. Sci.}, 104:424--428, 2007.

\bibitem{krishna_83}
S.~Krishnamurthy, R.~Bansil, and J.~Wiafe-Akenten.
\newblock Low-frequency raman spectrum of supercooled water.
\newblock {\em J. Chem. Phys.}, 79:5863--5870, 1983.

\bibitem{aliotta_86}
F.~Aliotta, C.~Vasi, G.~Maisano, D.~Majolino, F.~Mallamace, and P.~Migliardo.
\newblock Role of h bond and cooperative effects in normal and supercooled
  water studied by anisotropic low frequency light scattering.
\newblock {\em J. Chem. Phys.}, 84:4731--4738, 1986.

\bibitem{walrafen_86}
G.E. Walrafen, M.R. Fisher, M.S. Hokmabadi, and W.-H. Yang.
\newblock Temperature depencence of the low- and high-frequency raman
  scattering from liquid water.
\newblock {\em J. Chem. Phys.}, 85:6970--6982, 1986.

\bibitem{rousset_90}
J.~L. Rousset, E.~Duval, and A.~Boukenter.
\newblock Dynamical structure of water: Low-frequency raman scattering from a
  disordered network and aggregates.
\newblock {\em J. Chem. Phys.}, 92:2150--2154, 1990.

\bibitem{mizoguchi_92}
K.~Mizoguchi, Y.~Hori, and Y.~Tominaga.
\newblock Study on dynamical structure in water and heavy water by
  low-frequency raman spectroscopy.
\newblock {\em J. Chem. Phys.}, 97:1961--1968, 1992.

\bibitem{sokolov_95}
A.P. Sokolov, J.~Hurst, and D.~Quitmann.
\newblock Dynamics of supercooled water: Mode-coupling theory approach.
\newblock {\em Phys. Rev. B}, 51:12865--12868, 1995.

\bibitem{torre_08}
Renato Torre.
\newblock {\em Time-Resolved Spectroscopy in Complex Liquids, an experimental
  perspective}.
\newblock Springer, New York, 2008.

\bibitem{torre_04}
R.~Torre, P.~Bartolini, and R.~Righini.
\newblock Structural relaxation in super-cooled water by time-resolved
  spectroscopy.
\newblock {\em Nature}, 428:296--298, 2004.

\bibitem{taschin_06}
A.~Taschin, P.~Bartolini, R.~Eramo, and R.~Torre.
\newblock Supercooled water relaxation dynamics probed with heterodyne
  transient grating experiments.
\newblock {\em Phys. Rev. E}, 74:031502, 2006.

\bibitem{bartolini_09}
P.~Bartolini, A.~Taschin, R.~Eramo, R.~Righini, and R.~Torre.
\newblock Optical kerr effect measurements on supercooled water: the
  experimental perspective.
\newblock {\em J. of Physics: Conf. Series}, 177:012009, 2009.

\bibitem{taschin_11}
A.~Taschin, R.~Cucini, P.~Bartolini, and R.~Torre.
\newblock Does there exist an anomalous sound dispersion in supercooled water?
\newblock {\em Philos. Mag.}, 91:1796--1800, 2011.

\bibitem{castner_95}
E.W. Castner, Y.J. Chang, Y.C. Chu, and G.E. Walrafen.
\newblock The intermolecular dynamics of liquid water.
\newblock {\em J. Chem. Phys.}, 102:653--659, 1995.

\bibitem{palese_96}
S.~Palese, S.~Mukamel, R.J.D. Miller, and W.T. Lotshaw.
\newblock Interrogation of vibrational structure and line broadening of liquid
  water by raman-induced kerr effect measurements within multimode brownian
  oscillator model.
\newblock {\em J. Phys. Chem.}, 100:10380--10388, 1996.

\bibitem{winkler_02}
K.~Winkler, J.~Lindner, and P.~Vohringer.
\newblock Low-frequency depolarized raman-spectral density of liquid water from
  femtosecond optical kerr-effect measurements: Lineshape analysis of
  restricted translational modes.
\newblock {\em Phys. Chem. Chem. Phys.}, 4:2144--2155, 2002.

\bibitem{ratajska_06}
B.~Ratajska-Gadomska, B.~Bialkowski, W.~Gadomski, and Cz. Radzewicz.
\newblock Ultrashort memory of the quasicrystalline order in water by optical
  kerr effect spectroscopy.
\newblock {\em Chem. Phys. Lett.}, 429:575--580, 2006.

\bibitem{sonoda_05}
M.T. Sonoda, S.M. Vechi, and M.S. Skaf.
\newblock A simulation study of the optical kerr effect in liquid water.
\newblock {\em Phys. Chem. Chem. Phys.}, 7:1176--1180, 2005.

\bibitem{righini_93}
R.~Righini.
\newblock Ultrafast optical kerr effect in liquids and solids.
\newblock {\em Science}, 262(5138):1386--1390, November 1993.

\bibitem{hunt_07}
Neil~T. Hunt, Andrew~A. Jaye, and Stephen~R. Meech.
\newblock Ultrafast dynamics in complex fluids observed through the ultrafast
  optically-heterodyne-detected optical-kerr-effect (ohd-oke).
\newblock {\em Phys. Chem. Chem. Phys.}, 9:2167--2180, 2007.

\bibitem{bartolini_08}
P.Bartolini, A.~Taschin, R.~Eramo, and R.~Torre.
\newblock {\em Optical Kerr Effect Experiments on Complex Liquids, A Direct
  Access to Fast Dynamic Processes}, chapter~2, pages 73--127.
\newblock Springer, New York, 2008.

\bibitem{bartolini_99}
P.~Bartolini, M.~Ricci, R.~Torre, R.~Righini, and I.~Santa.
\newblock Diffusive and oscillatory dynamics of liquid iodobenzene measured by
  femtosecond optical kerr effect.
\newblock {\em J. Chem. Phys.}, 110:8653--8662, 1999.

\bibitem{torre_98}
R.~Torre, P.~Bartolini, and R.M. Pick.
\newblock Time-resolved optical kerr effect in a fragile glass-forming liquid,
  salol.
\newblock {\em Phys. Rev. E}, 57:1912--1920, 1998.

\bibitem{torre_00}
R.~Torre, P.~Bartolini, M.~Ricci, and R.M. Pick.
\newblock Time-resolved optical kerr effect on a fragile glass-forming liquid:
  test of different mode-coupling theory aspects.
\newblock {\em Europhys. Lett.}, 52:324--329, 2000.

\bibitem{prevosto_02}
D.~Prevosto, P.~Bartolini, R.~Torre, M.~Ricci, A.~Taschin, S.~Capaccioli,
  M.~Lucchesi., and P.Rolla.
\newblock Relaxation processes in a epoxy resin studied by time resolved
  optical kerr effect.
\newblock {\em Phys. Rev. E}, 66:11502, 2002.

\bibitem{ricci_02}
M.~Ricci, P.Bartolini, and R.~Torre.
\newblock Fast dynamics of a fragile glass-former by time resolved
  spectroscopy.
\newblock {\em Philos. Mag. B}, 82:541--551, 2002.

\bibitem{ricci_04}
M.~Ricci, S.~Wiebel, P.~Bartolini, A.~Taschin, and R.~Torre.
\newblock Time-resolved optical kerr effect experiments on supercooled benzene
  and test of mode-coupling theory.
\newblock {\em Philos. Mag.}, 84:1491--1499, 2004.

\bibitem{fecko_02}
C.J. Fecko, J.D. Eaves, and A.. Tokmakoff.
\newblock Isotropic and anisotropic raman scattering from molecular liquids
  measured by spatially masked optical kerr effect spectroscopy.
\newblock {\em J. Chem. Phys.}, 117(3):1139--1154, 2002.

\bibitem{kinoshita_95}
S.~Kinoshita, Y.~Kai, M.~Yamaguchi, and T.~Yagi.
\newblock Direct comparison between ultrafast optical kerr effect and
  high-resolution light scattering spectroscopy.
\newblock {\em Phys. Rev. Lett.}, 75:148--151, 1995.

\bibitem{skaf_05}
M.S. Skaf and M.T. Sonoda.
\newblock Optical kerr effect in supercooled water.
\newblock {\em Phys. Rev. Lett.}, 94:137802, 2005.

\bibitem{padro_04}
J.~A. Padr\'o and J.~Mart\'i.
\newblock Response to ‘‘comment on ‘an interpretation of the low-frequency
  spectrum of liquid water’ ’’j. chem. phys. 118, 452 (2003)".
\newblock {\em J. Chem. Phys.}, 120:1659--1660, 2004.

\bibitem{desantis_04}
A.~DeSantis, A.~Ercoli, and D.~Rocca.
\newblock Comment on ‘‘an interpretation of the low-frequency spectrum of
  liquid water’’ j. chem. phys. 118, 452 (2003).
\newblock {\em J. Chem. Phys.}, 120:1657--1658, 2004.

\bibitem{cho_94}
Minhaeng Cho, Graham~R. Fleming, Shinji Saito, Iwao Ohmine, and Richard~M.
  Stratt.
\newblock Instantaneous normal mode analysis of liquid water.
\newblock {\em J. Chem. Phys.}, 100(9):6672--6683, 1994.

\bibitem{huang_10}
Congcong Huang, T.M. Weiss, D.~Nordlund, K.T. Wikfeldt, L.G.M. Pettersson, and
  A.~Nilsson.
\newblock Increasing correlation length in bulk supercooled h2o, d2o, and nacl
  solution determined from small angle x-ray scattering.
\newblock {\em J. Chem. Phys.}, 133(13):134504, October 2010.

\bibitem{goetze_92}
W.~G\"otze and L.~Sj\"ogren.
\newblock Relaxation processes in supercooled liquids.
\newblock {\em Rep. on Progress in Physics}, 55:241--376, 1992.

\bibitem{goetze_00b}
W.~G\"otze and T.~Voigtmann.
\newblock Universal and non-universal features of glassy relaxation in
  propylene carbonate.
\newblock {\em Phys. Rev. E}, 61:4133--4147, 2000.

\bibitem{goetze_04}
W.~G\"otze and M.~Sperl.
\newblock Nearly-logarithmic decay of correlations in glass-forming liquids.
\newblock {\em Phys. Rev. Lett.}, 92:105701, 2004.

\bibitem{goetze_09}
W.~G\"otze.
\newblock {\em Complex Dynamics of Glass-Forming Liquids, a mode-coupling
  theory}.
\newblock Oxford: University Press, 2009.

\bibitem{suzuki_00}
Y.~Suzuki, Y.~Takasaki, Y.~Tominaga, and O.~Mishima.
\newblock Low-frequency raman spectra of amorphous ices.
\newblock {\em Chem. Phys. Lett.}, 319:81--84, 2000.

\bibitem{nilsson_12}
Anders Nilsson, Congcong Huang, and Lars~G.M. Pettersson.
\newblock Fluctuations in ambient water.
\newblock {\em J. Mol. Liq.}, 176:2--16, December 2012.

\bibitem{overduin_12}
S.~D. Overduin and G.~N. Patey.
\newblock Understanding the structure factor and isothermal compressibility of
  ambient water in terms of local structural environments.
\newblock {\em J. Phys. Chem. B}, 116(39):12014--12020, October 2012.

\bibitem{giraud_03}
G.~Giraud, C.M. Gordon, I.R. Dunkin, and K.~Wynne.
\newblock The effects of anion and cation substitution on the ultrafast solvent
  dynamics of ionic liquids: A time-resolved optical kerr-effect spectroscopic
  study.
\newblock {\em J. Chem. Phys.}, 119:464--477, 2003.

\bibitem{bartolini_07}
P.~Bartolini, R.~Eramo, A.~Taschin, M.~De Pas, and R.~Torre.
\newblock A real-time acquisition system for pump-probe spectroscopy.
\newblock {\em Philos. Mag.}, 87:731--740, 2007.

\bibitem{hellwarth_77}
R.~W. Hellwarth.
\newblock Third-order optical susceptibilities of liquids and solids.
\newblock {\em Prog. Quant. Electr.}, 5:1--68, 1977.

\bibitem{mcmorrow_88}
D.~McMorrow, W.T. Lotshaw, and G.A. Kenney-Wallace.
\newblock Femtosecond optical kerr studies on the origin of the nonlinear
  responses in simple liquids.
\newblock {\em IEEE J.Quantum Electron.}, QE-24:443--454, 1988.

\bibitem{torre_93}
R.~Torre, I.~Santa, and R.~Righini.
\newblock Pre-transitional effects in the liquid--plastic phase transition of
  p-terphenyl.
\newblock {\em Chem. Phys. Lett.}, 212:90--95, 1993.

\bibitem{alba_95}
C.~Alba-Simionesco and M.~Krauzman.
\newblock Low frequency raman spectroscopy of supercooled fragile liquids
  analyzed with schematic mode coupling models.
\newblock {\em J. Chem. Phys.}, 102(16):6574--6585, 1995.

\bibitem{krakoviack_02}
V.~Krakoviack and C.~Alba-Simionesco.
\newblock What can be learned from the schematic mode-coupling approach to experimental data?
\newblock {\em J. Chem. Phys.}, 117(5):2161--2171, 2002.

\bibitem{wiebel_02}
Sabine Wiebel and Joachim Wuttke.
\newblock Structural relaxation and mode coupling in a non-glassforming liquid:
  depolarized light scattering in benzene.
\newblock {\em New J. of Phys.}, 4:1--17, 2002.

\end{thebibliography}

\break
\newpage

%
%
%

\end{document}